%
%
%
%
%
%
%
%
\def\standardrisposta{s }\def\reducedrisposta{r }
\def\mplarisposta{mpla }\def\zerorisposta{z }
\def\doublerisposta{d }\def\cartarisposta{e }\def\amsrisposta{y }
\newcount\ingrandimento \newcount\sinnota \newcount\dimnota
\newcount\unoduecol \newdimen\collhsize \newdimen\tothsize
\newdimen\fullhsize \newcount\controllorisposta \sinnota=1
\newskip\infralinea  \global\controllorisposta=0
%
%
%
%
\def\risposta{s } 
\def\srisposta{e }
\def\arisposta{y }
\ifx\risposta\standardrisposta \ingrandimento=1200
\message {>> This will come out UNREDUCED << }
\dimnota=2 \unoduecol=1 \global\controllorisposta=1 \fi
\ifx\risposta\reducedrisposta \ingrandimento=1095 \dimnota=1
\unoduecol=1  \global\controllorisposta=1
\message {>> This will come out REDUCED << } \fi
\ifx\risposta\doublerisposta \ingrandimento=1000 \dimnota=2
\unoduecol=2   \message {>> You must print this in
LANDSCAPE orientation << } \global\controllorisposta=1 \fi
\ifx\risposta\mplarisposta \ingrandimento=1000 \dimnota=1
\message {>> Mod. Phys. Lett. A format << }
\unoduecol=1 \global\controllorisposta=1 \fi
\ifx\risposta\zerorisposta \ingrandimento=1000 \dimnota=2
\message {>> Zero Magnification format << }
\unoduecol=1 \global\controllorisposta=1 \fi
\ifnum\controllorisposta=0  \ingrandimento=1200
\message {>>> ERROR IN INPUT, I ASSUME STANDARD
UNREDUCED FORMAT <<< }  \dimnota=2 \unoduecol=1 \fi
\magnification=\ingrandimento
%
%
%
%
\newdimen\eucolumnsize \newdimen\eudoublehsize \newdimen\eudoublevsize
\newdimen\uscolumnsize \newdimen\usdoublehsize \newdimen\usdoublevsize
\newdimen\eusinglehsize \newdimen\eusinglevsize \newdimen\ussinglehsize
\newskip\standardbaselineskip \newdimen\ussinglevsize
\newskip\reducedbaselineskip \newskip\doublebaselineskip
\eucolumnsize=12.0truecm    
\eudoublehsize=25.5truecm   
\eudoublevsize=6.7truein    
\uscolumnsize=4.4truein     
\usdoublehsize=9.4truein    
\usdoublevsize=6.8truein    
\eusinglehsize=6.5truein    
\eusinglevsize=24truecm     
\ussinglehsize=6.5truein    
\ussinglevsize=8.9truein    
\standardbaselineskip=16pt plus.2pt  
\reducedbaselineskip=14pt plus.2pt   
\doublebaselineskip=12pt plus.2pt    
%
%
\def\Portoffset{}
\def\Landoffset{\voffset=-.2truein}
\ifx\risposta\mplarisposta \def\Portoffset{\hoffset=1.8truecm} \fi
%
%
\def\Landspec{}
\tolerance=10000
\parskip=0pt plus2pt  \leftskip=0pt \rightskip=0pt
%
%
\ifx\risposta\standardrisposta \infralinea=\standardbaselineskip \fi
\ifx\risposta\reducedrisposta  \infralinea=\reducedbaselineskip \fi
\ifx\risposta\doublerisposta   \infralinea=\doublebaselineskip \fi
\ifx\risposta\mplarisposta     \infralinea=13pt \fi
\ifx\risposta\zerorisposta     \infralinea=12pt plus.2pt\fi
\ifnum\controllorisposta=0    \infralinea=\standardbaselineskip \fi
\ifx\risposta\doublerisposta   \Landoffset \else \Portoffset \fi
\ifx\risposta\doublerisposta \ifx\srisposta\cartarisposta
\tothsize=\eudoublehsize \collhsize=\eucolumnsize
\vsize=\eudoublevsize  \else  \tothsize=\usdoublehsize
\collhsize=\uscolumnsize \vsize=\usdoublevsize \fi \else
\ifx\srisposta\cartarisposta \tothsize=\eusinglehsize
\vsize=\eusinglevsize \else  \tothsize=\ussinglehsize
\vsize=\ussinglevsize \fi \collhsize=4.4truein \fi
\ifx\risposta\mplarisposta \tothsize=5.0truein
\vsize=7.8truein \collhsize=4.4truein \fi
%
%
%
%
\newcount\contaeuler \newcount\contacyrill \newcount\contaams
\font\ninerm=cmr9  \font\eightrm=cmr8  \font\sixrm=cmr6
\font\ninei=cmmi9  \font\eighti=cmmi8  \font\sixi=cmmi6
\font\ninesy=cmsy9  \font\eightsy=cmsy8  \font\sixsy=cmsy6
\font\ninebf=cmbx9  \font\eightbf=cmbx8  \font\sixbf=cmbx6
\font\ninett=cmtt9  \font\eighttt=cmtt8  \font\nineit=cmti9
\font\eightit=cmti8 \font\ninesl=cmsl9  \font\eightsl=cmsl8
\skewchar\ninei='177 \skewchar\eighti='177 \skewchar\sixi='177
\skewchar\ninesy='60 \skewchar\eightsy='60 \skewchar\sixsy='60
\hyphenchar\ninett=-1 \hyphenchar\eighttt=-1 \hyphenchar\tentt=-1
%
\font\tencmmib=cmmib10  \newfam\cmmibfam  \skewchar\tencmmib='177
\font\tencmbsy=cmbsy10  \newfam\cmbsyfam  \skewchar\tencmbsy='60
\def\scaps{\cmcsc}                 
\font\tencmcsc=cmcsc10  \newfam\cmcscfam
\ifnum\ingrandimento=1095

\font\capsone=cmcsc10 at 10.95pt 

\else

\font\capsone=cmcsc10 at 12pt 
\fi

\def\ttaarr{\bf}                
\def\ppaarr{\sl}                

%
%
%
\newfam\eufmfam \newfam\msamfam \newfam\msbmfam \newfam\eufbfam
\def\Loadeulerfonts{\global\contaeuler=1 \ifx\arisposta\amsrisposta
\font\teneufm=eufm10              
\font\eighteufm=eufm8 \font\nineeufm=eufm9 \font\sixeufm=eufm6
\font\seveneufm=eufm7  \font\fiveeufm=eufm5
\font\teneufb=eufb10              
\font\eighteufb=eufb8 \font\nineeufb=eufb9 \font\sixeufb=eufb6
\font\seveneufb=eufb7  \font\fiveeufb=eufb5
\font\teneurm=eurm10              
\font\eighteurm=eurm8 \font\nineeurm=eurm9
\font\teneurb=eurb10              
\font\eighteurb=eurb8 \font\nineeurb=eurb9
\font\teneusm=eusm10              
\font\eighteusm=eusm8 \font\nineeusm=eusm9
\font\teneusb=eusb10              
\font\eighteusb=eusb8 \font\nineeusb=eusb9
\else \def\eufm{\tt} \def\eufb{\tt} \def\eurm{\tt} \def\eurb{\tt}
\def\eusm{\tt} \def\eusb{\tt}    \fi}

\def\loadamsmath{\global\contaams=1 \ifx\arisposta\amsrisposta
\font\tenmsam=msam10 \font\ninemsam=msam9 \font\eightmsam=msam8
\font\sevenmsam=msam7 \font\sixmsam=msam6 \font\fivemsam=msam5
\font\tenmsbm=msbm10 \font\ninemsbm=msbm9 \font\eightmsbm=msbm8
\font\sevenmsbm=msbm7 \font\sixmsbm=msbm6 \font\fivemsbm=msbm5
\else \def\msbm{\bf} \fi \def\Bbb{\msbm} \def\symbl{\msam} \tenpoint}
\def\loadcyrill{\global\contacyrill=1 \ifx\arisposta\amsrisposta
\font\tenwncyr=wncyr10 \font\ninewncyr=wncyr9 \font\eightwncyr=wncyr8
\font\tenwncyb=wncyr10 \font\ninewncyb=wncyr9 \font\eightwncyb=wncyr8
\font\tenwncyi=wncyr10 \font\ninewncyi=wncyr9 \font\eightwncyi=wncyr8
\else \def\cyrill{\sl} \def\cyrilb{\sl} \def\cyrili{\sl} \fi\tenpoint}
\ifx\arisposta\amsrisposta
\font\sevenex=cmex7               
\font\eightex=cmex8  \font\nineex=cmex9
\font\ninecmmib=cmmib9   \font\eightcmmib=cmmib8
\font\sevencmmib=cmmib7 \font\sixcmmib=cmmib6
\font\fivecmmib=cmmib5   \skewchar\ninecmmib='177
\skewchar\eightcmmib='177  \skewchar\sevencmmib='177
\skewchar\sixcmmib='177   \skewchar\fivecmmib='177
%
%
%
\def\ninecmbsy{\tencmbsy}
\def\eightcmbsy{\tencmbsy}
\def\sevencmbsy{\tencmbsy}
\def\sixcmbsy{\tencmbsy}
\def\fivecmbsy{\tencmbsy}
\font\ninecmcsc=cmcsc9    \font\eightcmcsc=cmcsc8     \else
\def\cmmib{\fam\cmmibfam\tencmmib}\textfont\cmmibfam=\tencmmib
\scriptfont\cmmibfam=\tencmmib \scriptscriptfont\cmmibfam=\tencmmib
\def\cmbsy{\fam\cmbsyfam\tencmbsy} \textfont\cmbsyfam=\tencmbsy
\scriptfont\cmbsyfam=\tencmbsy \scriptscriptfont\cmbsyfam=\tencmbsy
\scriptfont\cmcscfam=\tencmcsc \scriptscriptfont\cmcscfam=\tencmcsc
\def\cmcsc{\fam\cmcscfam\tencmcsc} \textfont\cmcscfam=\tencmcsc \fi
\catcode`@=11
\newskip\ttglue
\gdef\tenpoint{\def\rm{\fam0\tenrm}
  \textfont0=\tenrm \scriptfont0=\sevenrm \scriptscriptfont0=\fiverm
  \textfont1=\teni \scriptfont1=\seveni \scriptscriptfont1=\fivei
  \textfont2=\tensy \scriptfont2=\sevensy \scriptscriptfont2=\fivesy
  \textfont3=\tenex \scriptfont3=\tenex \scriptscriptfont3=\tenex
  \def\mcal{\fam2 \tensy}  \def\mmit{\fam1 \teni}
  \textfont\itfam=\tenit \def\it{\fam\itfam\tenit}
  \textfont\slfam=\tensl \def\sl{\fam\slfam\tensl}
  \textfont\ttfam=\tentt \scriptfont\ttfam=\eighttt
  \scriptscriptfont\ttfam=\eighttt  \def\tt{\fam\ttfam\tentt}
  \textfont\bffam=\tenbf \scriptfont\bffam=\sevenbf
  \scriptscriptfont\bffam=\fivebf \def\bf{\fam\bffam\tenbf}
     \ifx\arisposta\amsrisposta    \ifnum\contaeuler=1
  \textfont\eufmfam=\teneufm \scriptfont\eufmfam=\seveneufm
  \scriptscriptfont\eufmfam=\fiveeufm \def\eufm{\fam\eufmfam\teneufm}
  \textfont\eufbfam=\teneufb \scriptfont\eufbfam=\seveneufb
  \scriptscriptfont\eufbfam=\fiveeufb \def\eufb{\fam\eufbfam\teneufb}
  \def\eurm{\teneurm} \def\eurb{\teneurb} \def\eusm{\teneusm}
  \def\eusb{\teneusb}    \fi    \ifnum\contaams=1
  \textfont\msamfam=\tenmsam \scriptfont\msamfam=\sevenmsam
  \scriptscriptfont\msamfam=\fivemsam \def\msam{\fam\msamfam\tenmsam}
  \textfont\msbmfam=\tenmsbm \scriptfont\msbmfam=\sevenmsbm
  \scriptscriptfont\msbmfam=\fivemsbm \def\msbm{\fam\msbmfam\tenmsbm}
     \fi      \ifnum\contacyrill=1     \def\cyrill{\tenwncyr}
  \def\cyrilb{\tenwncyb}  \def\cyrili{\tenwncyi}         \fi
  \textfont3=\tenex \scriptfont3=\sevenex \scriptscriptfont3=\sevenex
  \def\cmmib{\fam\cmmibfam\tencmmib} \scriptfont\cmmibfam=\sevencmmib
  \textfont\cmmibfam=\tencmmib  \scriptscriptfont\cmmibfam=\fivecmmib
  \def\cmbsy{\fam\cmbsyfam\tencmbsy} \scriptfont\cmbsyfam=\sevencmbsy
  \textfont\cmbsyfam=\tencmbsy  \scriptscriptfont\cmbsyfam=\fivecmbsy
  \def\cmcsc{\fam\cmcscfam\tencmcsc} \scriptfont\cmcscfam=\eightcmcsc
  \textfont\cmcscfam=\tencmcsc \scriptscriptfont\cmcscfam=\eightcmcsc
     \fi            \tt \ttglue=.5em plus.25em minus.15em
  \normalbaselineskip=12pt
  \setbox\strutbox=\hbox{\vrule height8.5pt depth3.5pt width0pt}
  \let\sc=\eightrm \let\big=\tenbig   \normalbaselines
  \baselineskip=\infralinea  \rm}
\gdef\ninepoint{\def\rm{\fam0\ninerm}
  \textfont0=\ninerm \scriptfont0=\sixrm \scriptscriptfont0=\fiverm
  \textfont1=\ninei \scriptfont1=\sixi \scriptscriptfont1=\fivei
  \textfont2=\ninesy \scriptfont2=\sixsy \scriptscriptfont2=\fivesy
  \textfont3=\tenex \scriptfont3=\tenex \scriptscriptfont3=\tenex
  \def\mcal{\fam2 \ninesy}  \def\mmit{\fam1 \ninei}
  \textfont\itfam=\nineit \def\it{\fam\itfam\nineit}
  \textfont\slfam=\ninesl \def\sl{\fam\slfam\ninesl}
  \textfont\ttfam=\ninett \scriptfont\ttfam=\eighttt
  \scriptscriptfont\ttfam=\eighttt \def\tt{\fam\ttfam\ninett}
  \textfont\bffam=\ninebf \scriptfont\bffam=\sixbf
  \scriptscriptfont\bffam=\fivebf \def\bf{\fam\bffam\ninebf}
     \ifx\arisposta\amsrisposta  \ifnum\contaeuler=1
  \textfont\eufmfam=\nineeufm \scriptfont\eufmfam=\sixeufm
  \scriptscriptfont\eufmfam=\fiveeufm \def\eufm{\fam\eufmfam\nineeufm}
  \textfont\eufbfam=\nineeufb \scriptfont\eufbfam=\sixeufb
  \scriptscriptfont\eufbfam=\fiveeufb \def\eufb{\fam\eufbfam\nineeufb}
  \def\eurm{\nineeurm} \def\eurb{\nineeurb} \def\eusm{\nineeusm}
  \def\eusb{\nineeusb}     \fi   \ifnum\contaams=1
  \textfont\msamfam=\ninemsam \scriptfont\msamfam=\sixmsam
  \scriptscriptfont\msamfam=\fivemsam \def\msam{\fam\msamfam\ninemsam}
  \textfont\msbmfam=\ninemsbm \scriptfont\msbmfam=\sixmsbm
  \scriptscriptfont\msbmfam=\fivemsbm \def\msbm{\fam\msbmfam\ninemsbm}
     \fi       \ifnum\contacyrill=1     \def\cyrill{\ninewncyr}
  \def\cyrilb{\ninewncyb}  \def\cyrili{\ninewncyi}         \fi
  \textfont3=\nineex \scriptfont3=\sevenex \scriptscriptfont3=\sevenex
  \def\cmmib{\fam\cmmibfam\ninecmmib}  \textfont\cmmibfam=\ninecmmib
  \scriptfont\cmmibfam=\sixcmmib \scriptscriptfont\cmmibfam=\fivecmmib
  \def\cmbsy{\fam\cmbsyfam\ninecmbsy}  \textfont\cmbsyfam=\ninecmbsy
  \scriptfont\cmbsyfam=\sixcmbsy \scriptscriptfont\cmbsyfam=\fivecmbsy
  \def\cmcsc{\fam\cmcscfam\ninecmcsc} \scriptfont\cmcscfam=\eightcmcsc
  \textfont\cmcscfam=\ninecmcsc \scriptscriptfont\cmcscfam=\eightcmcsc
     \fi            \tt \ttglue=.5em plus.25em minus.15em
  \normalbaselineskip=11pt
  \setbox\strutbox=\hbox{\vrule height8pt depth3pt width0pt}
  \let\sc=\sevenrm \let\big=\ninebig \normalbaselines\rm}
\gdef\eightpoint{\def\rm{\fam0\eightrm}
  \textfont0=\eightrm \scriptfont0=\sixrm \scriptscriptfont0=\fiverm
  \textfont1=\eighti \scriptfont1=\sixi \scriptscriptfont1=\fivei
  \textfont2=\eightsy \scriptfont2=\sixsy \scriptscriptfont2=\fivesy
  \textfont3=\tenex \scriptfont3=\tenex \scriptscriptfont3=\tenex
  \def\mcal{\fam2 \eightsy}  \def\mmit{\fam1 \eighti}
  \textfont\itfam=\eightit \def\it{\fam\itfam\eightit}
  \textfont\slfam=\eightsl \def\sl{\fam\slfam\eightsl}
  \textfont\ttfam=\eighttt \scriptfont\ttfam=\eighttt
  \scriptscriptfont\ttfam=\eighttt \def\tt{\fam\ttfam\eighttt}
  \textfont\bffam=\eightbf \scriptfont\bffam=\sixbf
  \scriptscriptfont\bffam=\fivebf \def\bf{\fam\bffam\eightbf}
     \ifx\arisposta\amsrisposta   \ifnum\contaeuler=1
  \textfont\eufmfam=\eighteufm \scriptfont\eufmfam=\sixeufm
  \scriptscriptfont\eufmfam=\fiveeufm \def\eufm{\fam\eufmfam\eighteufm}
  \textfont\eufbfam=\eighteufb \scriptfont\eufbfam=\sixeufb
  \scriptscriptfont\eufbfam=\fiveeufb \def\eufb{\fam\eufbfam\eighteufb}
  \def\eurm{\eighteurm} \def\eurb{\eighteurb} \def\eusm{\eighteusm}
  \def\eusb{\eighteusb}       \fi    \ifnum\contaams=1
  \textfont\msamfam=\eightmsam \scriptfont\msamfam=\sixmsam
  \scriptscriptfont\msamfam=\fivemsam \def\msam{\fam\msamfam\eightmsam}
  \textfont\msbmfam=\eightmsbm \scriptfont\msbmfam=\sixmsbm
  \scriptscriptfont\msbmfam=\fivemsbm \def\msbm{\fam\msbmfam\eightmsbm}
     \fi       \ifnum\contacyrill=1     \def\cyrill{\eightwncyr}
  \def\cyrilb{\eightwncyb}  \def\cyrili{\eightwncyi}         \fi
  \textfont3=\eightex \scriptfont3=\sevenex \scriptscriptfont3=\sevenex
  \def\cmmib{\fam\cmmibfam\eightcmmib}  \textfont\cmmibfam=\eightcmmib
  \scriptfont\cmmibfam=\sixcmmib \scriptscriptfont\cmmibfam=\fivecmmib
  \def\cmbsy{\fam\cmbsyfam\eightcmbsy}  \textfont\cmbsyfam=\eightcmbsy
  \scriptfont\cmbsyfam=\sixcmbsy \scriptscriptfont\cmbsyfam=\fivecmbsy
  \def\cmcsc{\fam\cmcscfam\eightcmcsc} \scriptfont\cmcscfam=\eightcmcsc
  \textfont\cmcscfam=\eightcmcsc \scriptscriptfont\cmcscfam=\eightcmcsc
     \fi             \tt \ttglue=.5em plus.25em minus.15em
  \normalbaselineskip=9pt
  \setbox\strutbox=\hbox{\vrule height7pt depth2pt width0pt}
  \let\sc=\sixrm \let\big=\eightbig \normalbaselines\rm }
\gdef\tenbig#1{{\hbox{$\left#1\vbox to8.5pt{}\right.\n@space$}}}
\gdef\ninebig#1{{\hbox{$\textfont0=\tenrm\textfont2=\tensy
   \left#1\vbox to7.25pt{}\right.\n@space$}}}
\gdef\eightbig#1{{\hbox{$\textfont0=\ninerm\textfont2=\ninesy
   \left#1\vbox to6.5pt{}\right.\n@space$}}}
\def\alternativefont#1#2{\ifx\arisposta\amsrisposta \relax \else
\xdef#1{#2} \fi}
\global\contaeuler=0 \global\contacyrill=0 \global\contaams=0
%
%
%
%
\newbox\fotlinebb \newbox\hedlinebb \newbox\leftcolumn
\gdef\makeheadline{\vbox to 0pt{\vskip-22.5pt
     \fullline{\vbox to8.5pt{}\the\headline}\vss}\nointerlineskip}
\gdef\makehedlinebb{\vbox to 0pt{\vskip-22.5pt
     \fullline{\vbox to8.5pt{}\copy\hedlinebb\hfil
     \line{\hfill\the\headline\hfill}}\vss} \nointerlineskip}
\gdef\makefootline{\baselineskip=24pt \fullline{\the\footline}}
\gdef\makefotlinebb{\baselineskip=24pt
    \fullline{\copy\fotlinebb\hfil\line{\hfill\the\footline\hfill}}}
\gdef\doubleformat{\shipout\vbox{\Landspec\makehedlinebb
     \fullline{\box\leftcolumn\hfil\columnbox}\makefotlinebb}
     \advancepageno}
\gdef\columnbox{\leftline{\pagebody}}
\gdef\line#1{\hbox to\hsize{\hskip\leftskip#1\hskip\rightskip}}
\gdef\fullline#1{\hbox to\fullhsize{\hskip\leftskip{#1}%
\hskip\rightskip}}
\gdef\footnote#1{\let\@sf=\empty
         \ifhmode\edef\#sf{\spacefactor=\the\spacefactor}\/\fi
         #1\@sf\vfootnote{#1}}
\gdef\vfootnote#1{\insert\footins\bgroup
         \ifnum\dimnota=1  \eightpoint\fi
         \ifnum\dimnota=2  \ninepoint\fi
         \ifnum\dimnota=0  \tenpoint\fi
         \interlinepenalty=\interfootnotelinepenalty
         \splittopskip=\ht\strutbox
         \splitmaxdepth=\dp\strutbox \floatingpenalty=20000
         \leftskip=\oldssposta \rightskip=\olddsposta
         \spaceskip=0pt \xspaceskip=0pt
         \ifnum\sinnota=0   \textindent{#1}\fi
         \ifnum\sinnota=1   \item{#1}\fi
         \footstrut\futurelet\next\fo@t}
\gdef\fo@t{\ifcat\bgroup\noexpand\next \let\next\f@@t
             \else\let\next\f@t\fi \next}
\gdef\f@@t{\bgroup\aftergroup\@foot\let\next}
\gdef\f@t#1{#1\@foot} \gdef\@foot{\strut\egroup}
\gdef\footstrut{\vbox to\splittopskip{}}
\skip\footins=\bigskipamount
\count\footins=1000  \dimen\footins=8in
\catcode`@=12
\tenpoint
\ifnum\unoduecol=1 \hsize=\tothsize   \fullhsize=\tothsize \fi
\ifnum\unoduecol=2 \hsize=\collhsize  \fullhsize=\tothsize \fi
\global\let\lrcol=L      \ifnum\unoduecol=1
\output{\plainoutput{\ifnum\tipbnota=2 \clearnmbnota\fi}} \fi
\ifnum\unoduecol=2 \output{\if L\lrcol
     \global\setbox\leftcolumn=\columnbox
     \global\setbox\fotlinebb=\line{\hfill\the\footline\hfill}
     \global\setbox\hedlinebb=\line{\hfill\the\headline\hfill}
     \advancepageno  \global\let\lrcol=R
     \else  \doubleformat \global\let\lrcol=L \fi
     \ifnum\outputpenalty>-20000 \else\dosupereject\fi
     \ifnum\tipbnota=2\clearnmbnota\fi }\fi
\def\ifdoublepage{\ifnum\unoduecol=2 }
\gdef\yespagenumbers{\footline={\hss\tenrm\folio\hss}}
\gdef\ciao{ \ifnum\fdefcontre=1 \endfdef\fi
     \par\vfill\supereject \ifnum\unoduecol=2
     \if R\lrcol  \headline={}\nopagenumbers\null\vfill\eject
     \fi\fi \end}

\newskip\olddsposta \newskip\oldssposta
\global\oldssposta=\leftskip \global\olddsposta=\rightskip

\def\filldots{\leaders\hbox to 1em{\hss.\hss}\hfill}
\def\inquadrb#1 {\vbox {\hrule  \hbox{\vrule \vbox {\vskip .2cm
    \hbox {\ #1\ } \vskip .2cm } \vrule  }  \hrule} }
 \def\newline{\hfil\break}
\def\jump{\vskip\baselineskip} \newskip\iinnffrr
\def\sjump{\iinnffrr=\baselineskip
          \divide\iinnffrr by 2 \vskip\iinnffrr}
\def\bjump{\vskip\baselineskip \vskip\baselineskip}
\newcount\nmbnota  \def\clearnmbnota{\global\nmbnota=0}
\newcount\tipbnota \def\letterfootnote{\global\tipbnota=1}

\def\note#1{\global\advance\nmbnota by 1 \ifnum\tipbnota=1
    \footnote{$^{\rm\nttlett}$}{#1} \else {\ifnum\tipbnota=2
    \footnote{$^{\nttsymb}$}{#1}
    \else\footnote{$^{\the\nmbnota}$}{#1}\fi}\fi}
\def\nttlett{\ifcase\nmbnota \or a\or b\or c\or d\or e\or f\or
g\or h\or i\or j\or k\or l\or m\or n\or o\or p\or q\or r\or
s\or t\or u\or v\or w\or y\or x\or z\fi}
\def\nttsymb{\ifcase\nmbnota \or\dag\or\sharp\or\ddag\or\star\or
\natural\or\flat\or\clubsuit\or\diamondsuit\or\heartsuit
\or\spadesuit\fi}   \clearnmbnota
\def\numberfootnote{\global\tipbnota=0} \numberfootnote
\def\setnote#1{\expandafter\xdef\csname#1\endcsname{
\ifnum\tipbnota=1 {\rm\nttlett} \else {\ifnum\tipbnota=2
{\nttsymb} \else \the\nmbnota\fi}\fi} }
\newcount\nbmfig  \def\clearnbmfig{\global\nbmfig=0}
\gdef\figure{\global\advance\nbmfig by 1
      {\rm fig. \the\nbmfig}}   \clearnbmfig
\def\setfig#1{\expandafter\xdef\csname#1\endcsname{fig. \the\nbmfig}}
 \def\endformula{\eqno\numero $$}
 \def\efr{\endformula}
\newcount\frmcount \def\clearfrmcount{\global\frmcount=0}
\def\numero{\global\advance\frmcount by 1   \ifnum\indappcount=0
  {\ifnum\cpcount <1 {\hbox{\rm (\the\frmcount )}}  \else
  {\hbox{\rm (\the\cpcount .\the\frmcount )}} \fi}  \else
  {\hbox{\rm (\applett .\the\frmcount )}} \fi}
\def\nameformula#1{\global\advance\frmcount by 1%
\ifnum\draftnum=0  {\ifnum\indappcount=0%
{\ifnum\cpcount<1\xdef\spzzttrra{(\the\frmcount )}%
\else\xdef\spzzttrra{(\the\cpcount .\the\frmcount )}\fi}%
\else\xdef\spzzttrra{(\applett .\the\frmcount )}\fi}%
\else\xdef\spzzttrra{(#1)}\fi%
\expandafter\xdef\csname#1\endcsname{\spzzttrra}
\eqno \hbox{\rm\spzzttrra} $$}
\def\nfr{\nameformula}    
\def\nameali#1{\global\advance\frmcount by 1%
\ifnum\draftnum=0  {\ifnum\indappcount=0%
{\ifnum\cpcount<1\xdef\spzzttrra{(\the\frmcount )}%
\else\xdef\spzzttrra{(\the\cpcount .\the\frmcount )}\fi}%
\else\xdef\spzzttrra{(\applett .\the\frmcount )}\fi}%
\else\xdef\spzzttrra{(#1)}\fi%
\expandafter\xdef\csname#1\endcsname{\spzzttrra}
  \hbox{\rm\spzzttrra} }      \clearfrmcount
\newcount\cpcount \def\clearcpcount{\global\cpcount=0}
\newcount\subcpcount \def\clearsubcpcount{\global\subcpcount=0}
\newcount\appcount \def\clearappcount{\global\appcount=0}
\newcount\indappcount \def\clearindappcount{\indappcount=0}
\newcount\sottoparcount 

\def\applett{\ifcase\appcount  \or {A}\or {B}\or {C}\or
{D}\or {E}\or {F}\or {G}\or {H}\or {I}\or {J}\or {K}\or {L}\or
{M}\or {N}\or {O}\or {P}\or {Q}\or {R}\or {S}\or {T}\or {U}\or
{V}\or {W}\or {X}\or {Y}\or {Z}\fi    \ifnum\appcount<0
\immediate\write16 {Panda ERROR - Appendix: counter "appcount"
out of range}\fi  \ifnum\appcount>26  \immediate\write16 {Panda
ERROR - Appendix: counter "appcount" out of range}\fi}
\clearappcount  \clearindappcount \newcount\connttrre
\def\clearconnttrre{\global\connttrre=0} \newcount\countref
\def\clearcountref{\global\countref=0} \clearcountref
\def\chapter#1{\global\advance\cpcount by 1 \clearfrmcount
                 \goodbreak\null\vbox{\jump\nobreak
                 \clearsubcpcount\clearindappcount
                 \itemitem{\ttaarr\the\cpcount .\qquad}{\ttaarr #1}
                 \par\nobreak\jump\sjump}\nobreak}
\def\section#1{\global\advance\subcpcount by 1 \goodbreak\null
               \vbox{\sjump\nobreak\ifnum\indappcount=0
                 {\ifnum\cpcount=0 {\itemitem{\ppaarr
               .\the\subcpcount\quad\enskip\ }{\ppaarr #1}\par} \else
                 {\itemitem{\ppaarr\the\cpcount .\the\subcpcount\quad
                  \enskip\ }{\ppaarr #1} \par}  \fi}
                \else{\itemitem{\ppaarr\applett .\the\subcpcount\quad
                 \enskip\ }{\ppaarr #1}\par}\fi\nobreak\jump}\nobreak}
\clearsubcpcount
\def\appendix#1{\global\advance\appcount by 1 \clearfrmcount
                  \goodbreak\null\vbox{\jump\nobreak
                  \global\advance\indappcount by 1 \clearsubcpcount
          \itemitem{ }{\hskip-40pt\ttaarr Appendix\ #1}
             \nobreak\jump\sjump}\nobreak}
\clearappcount \clearindappcount
\def\references{\goodbreak\null\vbox{\jump\nobreak
   \itemitem{}{\ttaarr References} \nobreak\jump\sjump}\nobreak}

\clearcpcount\clearcountref

\def\setchap#1{\ifnum\indappcount=0{\ifnum\subcpcount=0%
\xdef\spzzttrra{\the\cpcount}%
\else\xdef\spzzttrra{\the\cpcount .\the\subcpcount}\fi}
\else{\ifnum\subcpcount=0 \xdef\spzzttrra{\applett}%
\else\xdef\spzzttrra{\applett .\the\subcpcount}\fi}\fi
\expandafter\xdef\csname#1\endcsname{\spzzttrra}}
\newcount\draftnum \newcount\ppora   \newcount\ppminuti
\global\ppora=\time   \global\ppminuti=\time
\global\divide\ppora by 60  \draftnum=\ppora
\multiply\draftnum by 60    \global\advance\ppminuti by -\draftnum
\def\droggi{\number\day /\number\month /\number\year\ \the\ppora
:\the\ppminuti}     \global\draftnum=0
\def\draftcomment#1{\ifnum\draftnum=0 \relax \else
{\ {\bf ***}\ #1\ {\bf ***}\ }\fi} 
%
%
\catcode`@=11
\gdef\Ref#1{\expandafter\ifx\csname @rrxx@#1\endcsname\relax%
{\global\advance\countref by 1    \ifnum\countref>200
\immediate\write16 {Panda ERROR - Ref: maximum number of references
exceeded}  \expandafter\xdef\csname @rrxx@#1\endcsname{0}\else
\expandafter\xdef\csname @rrxx@#1\endcsname{\the\countref}\fi}\fi
\ifnum\draftnum=0 \csname @rrxx@#1\endcsname \else#1\fi}
\gdef\beginref{\ifnum\draftnum=0  \gdef\Rref{\fairef}
\gdef\endref{\scriviref} \else\relax\fi
\ifx\risposta\mplarisposta \ninepoint \fi
\parskip 2pt plus.2pt \baselineskip=12pt}
\def\Reflab#1{[#1]} \gdef\Rref#1#2{\item{\Reflab{#1}}{#2}}
\gdef\endref{\relax}  \newcount\conttemp
\gdef\fairef#1#2{\expandafter\ifx\csname @rrxx@#1\endcsname\relax
{\global\conttemp=0 \immediate\write16 {Panda ERROR - Ref: reference
[#1] undefined}} \else
{\global\conttemp=\csname @rrxx@#1\endcsname } \fi
\global\advance\conttemp by 50  \global\setbox\conttemp=\hbox{#2} }
\gdef\scriviref{\clearconnttrre\conttemp=50
\loop\ifnum\connttrre<\countref \advance\conttemp by 1
\advance\connttrre by 1
\item{\Reflab{\the\connttrre}}{\unhcopy\conttemp} \repeat}
\clearcountref \clearconnttrre
\catcode`@=12
\ifx\risposta\mplarisposta \def\Reflab#1{#1.} \letterfootnote \fi

\def\slashchar#1{\setbox0=\hbox{$#1$} \dimen0=\wd0
     \setbox1=\hbox{/} \dimen1=\wd1 \ifdim\dimen0>\dimen1
      \rlap{\hbox to \dimen0{\hfil/\hfil}} #1 \else
      \rlap{\hbox to \dimen1{\hfil$#1$\hfil}} / \fi}
\ifx\oldchi\undefined \let\oldchi=\chi
  \def\cchi{{\raise 1pt\hbox{$\oldchi$}}} \let\chi=\cchi \fi
  
\def\del{\partial}   

\def\frac#1#2{{\textstyle{#1 \over #2}}}

\def\half{\ifinner {\scriptstyle {1 \over 2}}\else {1 \over 2} \fi}

\def\simge{\rlap{\raise 2pt \hbox{$>$}}{\lower 2pt \hbox{$\sim$}}}
\def\simle{\rlap{\raise 2pt \hbox{$<$}}{\lower 2pt \hbox{$\sim$}}}

\def\vbig#1#2{{\vbigd@men=#2\divide\vbigd@men by 2%
\hbox{$\left#1\vbox to \vbigd@men{}\right.\n@space$}}}

%
%
\newcount\fdefcontre \newcount\fdefcount \newcount\indcount
\newread\filefdef  \newread\fileftmp  \newwrite\filefdef
\newwrite\fileftmp     \def\strip#1*.A {#1}
\def\futuredef#1{\beginfdef
\expandafter\ifx\csname#1\endcsname\relax%
{\immediate\write\fileftmp {#1*.A}
\immediate\write16 {Panda Warning - fdef: macro "#1" on page
\the\pageno \space undefined}
\ifnum\draftnum=0 \expandafter\xdef\csname#1\endcsname{(?)}
\else \expandafter\xdef\csname#1\endcsname{(#1)} \fi
\global\advance\fdefcount by 1}\fi   \csname#1\endcsname}

\def\beginfdef{\ifnum\fdefcontre=0
\immediate\openin\filefdef \jobname.fdef
\immediate\openout\fileftmp \jobname.ftmp
\global\fdefcontre=1  \ifeof\filefdef \immediate\write16 {Panda
WARNING - fdef: file \jobname.fdef not found, run TeX again}
\else \immediate\read\filefdef to\spzzttrra
\global\advance\fdefcount by \spzzttrra
\indcount=0      \loop\ifnum\indcount<\fdefcount
\advance\indcount by 1   \immediate\read\filefdef to\spezttrra
\immediate\read\filefdef to\sppzttrra
\edef\spzzttrra{\expandafter\strip\spezttrra}
\immediate\write\fileftmp {\spzzttrra *.A}
\expandafter\xdef\csname\spzzttrra\endcsname{\sppzttrra}
\repeat \fi \immediate\closein\filefdef \fi}
\def\endfdef{\immediate\closeout\fileftmp   \ifnum\fdefcount>0
\immediate\openin\fileftmp \jobname.ftmp
\immediate\openout\filefdef \jobname.fdef
\immediate\write\filefdef {\the\fdefcount}   \indcount=0
\loop\ifnum\indcount<\fdefcount    \advance\indcount by 1
\immediate\read\fileftmp to\spezttrra
\edef\spzzttrra{\expandafter\strip\spezttrra}
\immediate\write\filefdef{\spzzttrra *.A}
\edef\spezttrra{\string{\csname\spzzttrra\endcsname\string}}
\iwritel\filefdef{\spezttrra}
\repeat  \immediate\closein\fileftmp \immediate\closeout\filefdef
\immediate\write16 {Panda Warning - fdef: Label(s) may have changed,
re-run TeX to get them right}\fi}
\def\iwritel#1#2{\newlinechar=-1
{\newlinechar=`\ \immediate\write#1{#2}}\newlinechar=-1}
\global\fdefcontre=0 \global\fdefcount=0 \global\indcount=0
%
%
\null
%
%
%
%

\loadamsmath
%

\def\A{{\rm A}}
\def\B{{\rm B}}
\def\C{{\rm C}}
\def\D{{\rm D}}

\def\X{{\rm X}}
\def\S{{\cal S}}
\def\delb{\bar \del}
\def\zb{\bar z}
\def\wb{\bar w} 
\def\L{{\cal L}}
\def\W{{\cal W}}
\def\Half{{\textstyle{1\over2}}}
\nopagenumbers{\baselineskip=12pt
\rightline{CERN-TH/96-224} 
\rightline{DAMTP/96-76} 
\rightline{ENSLAPP-A-612/96} 
\rightline{\hfill hep-th/9608190}
\ifdoublepage \bjump\bjump\bjump\bjump\else\vfill\fi
\centerline{\capsone INTEGRABILITY vs.~SUPERSYMMETRY} 
\bjump\bjump
\centerline{\scaps Jonathan M.~Evans\note{
Supported by a PPARC Advanced Fellowship}
}
\sjump
\centerline{\sl Theoretical Physics Division, CERN} 
\centerline{\sl CH-1211, Geneva 23, Switzerland}
\centerline{\sl and }
\centerline{\sl DAMTP, University of Cambridge}
\centerline{\sl Silver Street, Cambridge CB3 9EW, UK} 
\centerline{\tt J.M.Evans@damtp.cam.ac.uk}
\sjump
\sjump
\centerline{\scaps Jens Ole Madsen}
\sjump
\centerline{\sl Laboratoire de Physique Th{\'e}orique}
\centerline{\sl ENSLAPP\note{
URA 14-36 du CNRS, associ{\'e}e {\`a} l'E.N.S. de Lyon et {\`a} 
l'Universit{\'e} de Savoie}, Groupe d'Annecy}
\centerline{\sl Chemin de Bellevue, F-74941 Annecy le Vieux, France} 
\centerline{\tt madsen@lapp.in2p3.fr}

\vfill
\ifnum\unoduecol=2 \eject\null\vfill\fi
\centerline{\capsone ABSTRACT}
\sjump
\noindent 
We investigate (1,0)-superconformal Toda theories based 
on simple Lie algebras and find 
that the classical integrability properties of
the underlying bosonic theories do not survive.
For several models based on algebras of low rank, we show 
explicitly that 
none of the conserved $\W$-algebra generators 
can be generalized to the supersymmetric case. 
Using these results we deduce that at least one 
$\W$-algebra generator fails to
generalize in any model based on a classical Lie algebra. 
This argument involves a method for relating the bosonic Toda 
theories and their conserved currents within each classical series.
We also scrutinize claims that the (1,0)-superconformal 
models actually admit (1,1) supersymmetry and find that they do not.
Our results are consistent with the belief that all integrable
Toda models with fermions arise from Lie superalgebras.
\sjump
\ifnum\unoduecol=2 \vfill\fi
\leftline{CERN-TH/96-224}
\leftline{DAMTP/96-76}
\leftline{ENSLAPP-A-612/96}
\leftline{August 1996}
\eject
\yespagenumbers\pageno=1

\chapter{Introduction}

Bosonic Toda theories [\Ref{LS}] have been studied for a number of years 
as important examples of 
integrable field theories in two dimensions; see
eg.~[\Ref{MS},\Ref{BS},\Ref{O`R}] 
for introductions to some aspects relevant to this paper.
The problem of incorporating fermions in the Toda construction has 
been considered by many authors 
$\vphantom{\hbox{
[\Ref{LLS},\Ref{O},\Ref{LSS},\Ref{LM},\Ref{EH},\Ref{NM},\Ref{IK},\Ref{WI},
\Ref{SHR},\Ref{ABL},\Ref{GPZ},\Ref{QI},\Ref{STS}]}}$
[5-17]
with much of the attention focused on finding 
supersymmetric models.
One conclusion which has emerged is that 
the bosonic Toda models based on simple
Lie algebras (or their Kac-Moody extensions) cannot be
supersymmetrized, except in the simplest case of Liouville (or 
sinh-Gordon) theory. More precisely, it is believed that there is 
no $N=1$ supersymmetric theory whose bosonic part is a 
Toda model based on a
{\it simple\/} Lie algebra of rank bigger than one.
To find integrable Toda theories with fermions, 
a more radical generalization 
of the bosonic construction is needed in which
the underlying Lie algebra is replaced with a Lie
superalgebra.  
Integrability of these models can be understood in a
uniform way by casting 
the field equations as a zero-curvature condition from which
one can extract conserved quantities.

The possibility of supersymmetrizing the bosonic Toda models was
reconsidered recently in an interesting paper by Papadopoulos
[\Ref{Pap}].
Working within the general framework of sigma-models with a potential term
[\Ref{HPT}],
he pointed out that the conformal Toda models 
based on simple Lie algebras
all admit
(1,0)-supersymmetric extensions.
This means that the theories in question possess 
a single conserved supercharge of definite
two-dimensional chirality, as opposed to $(1,1)$ 
supersymmetry, which would involve supercharges of both chiralities.
It is this second possibility which we referred to above simply as 
$N=1$ supersymmetry and which is believed to be ruled out.  

Since the new (1,0) models are supersymmetric extensions of
integrable theories it is tempting to think that they too must be
integrable. However, it seems that there is no obvious way to write
the equations of motion as a zero-curvature condition of the type
that guarantees integrability in the bosonic cases.
Another puzzle emerges when one
examines the known properties of extended chiral algebras appearing in
conformal field theory [\Ref{BS}].
The conserved quantities in the bosonic Toda models 
form $\W$-algebras---the models based on $\A_n$, for
example, realize the algebras usually referred to as
$\W_{n+1}$. The (1,0) models, if similarly
integrable, would be expected to contain supersymmetric generalizations of 
these $\W$-algebras 
which should exist at both the classical and quantum levels.
Minimal supersymmetric versions of the $\W_n$ algebras have been constructed
[\Ref{SW}], but
they are ``exotic'' in the terminology of [\Ref{BS}]
(``non-deformable'' in the
terminology of [\Ref{BW}]) meaning that they are associative for only a 
finite set of values of the central charge and therefore have no
classical limit.

These observations suggest that 
the (1,0)-superconformal models are either not integrable, 
or else they must contain some much
larger and more complicated chiral algebra structure in which the
bosonic $\W$-algebras are embedded. 
In this paper we aim to settle this issue
by showing that the process of
supersymmetrizing destroys the $\W$-symmetry present in 
the bosonic models.
These are the first examples we know of in which 
supersymmetry has this destructive effect on integrability. 
Our arguments are based on knowledge of the bosonic Toda theories
together with explicit calculations for (1,0) models corresponding to
low rank algebras. We then show how these facts can be put together to
draw conclusions about the (1,0) models based on general classical
algebras.

\chapter{Bosonic conformal Toda models} 

We recall some details of bosonic Toda theories that will be needed
later. Let $\X_n$ be any simple, finite-dimensional Lie algebra of
rank $n$, and $\alpha_i$ ($i=1, \ldots , n$) a set of simple roots.
The $\X_n$ Toda model can be defined in two-dimensional Minkowski
space by a Lagrangian
$$
L = \del \phi \cdot \delb \phi 
- \sum_i \exp( \alpha_i \cdot \phi )
\nfr{lag}
where $\phi (z ,\zb)$ is a field 
taking values in the Cartan subalgebra of $\X_n$ and a dot denotes the
invariant inner-product.\note{It is customary to include an explicit  
dimensionless coupling $\beta$ but we have chosen to absorb this 
in the field $\phi$. The normalization of
the kinetic terms is also non-standard, to allow a simpler comparison
with the supersymmetric models of the next section.} 
We use light-cone coordinates $z = \half (t-x)$ and $\bar z =
\half(t+x)$ and we
shall refer to quantities depending solely on $z$ or $\bar z$ as 
holomorphic or anti-holomorphic respectively. 

The Lagrangian above is invariant under the classical conformal 
transformations 
$$
\phi (z , \zb) \rightarrow \phi( w , \wb ) + 
\rho \log ({\del w} {\delb \wb}) 
\nfr{conf}
where $w(z)$, $\wb (\zb)$ are independent real-analytic functions and
the vector $\rho$ is defined by the property 
$\rho \cdot \alpha_i = 1$ for each simple root.
The corresponding energy-momentum tensor is traceless, with holomorphic and
anti-holomorphic components 
$$
T = \Half \del \phi \cdot \del \phi - \rho \cdot
\del^2 \phi \, , \qquad
\bar T = \Half \delb \phi \cdot \delb \phi - \rho \cdot \delb^2 \phi
\, .
\nfr{bosem}
Recall that a quantity $U(z,\zb)$ is said to have 
scaling-dimensions $(h , \bar h)$ if under the transformations $w =
\mu z$ and $\bar w =
\bar \mu \zb$ it behaves as 
$
U(z , \zb) \rightarrow \mu^h {\bar \mu}^{\bar h} U( \mu z , \bar
\mu \zb) 
$ 
where $\mu $, $\bar \mu$ are positive real numbers. 
We shall refer to objects with scaling dimension $(q,0)$ or $(0,q)$
as being of spin $q$. 
Notice that the field 
$\phi$ does not have definite scaling-dimensions, but that $\del \phi$
and $\delb \phi$ have scaling-dimensions (1,0) and (0,1) respectively.
In addition we can 
construct the potential-like
terms 
$\exp (\kappa \alpha_i \cdot \phi )$ which have 
scaling-dimensions $(\kappa , \kappa)$.

The $\A_1$ Toda model is just the Liouville theory, for which 
conformal invariance alone is sufficient to
establish integrability.
For rank $n > 1$, however, integrability of the
$\X_n$ model depends on the presence of higher-spin conserved quantities
which, together with the energy-momentum tensor, form a $\W$-algebra. 
The simplest example is the $\A_2$ Toda theory 
in which there is a spin-3 holomorphic current $W$ in addition
to the spin-2 current $T$. Explicitly, 
$$\eqalign{
T & = {1 \over 3} ( {(\del \phi_1)}^2 + {(\del \phi_2)}^2 
+ \del \phi_1 \del \phi_2 ) 
- (\del^2 \phi_1 + \del ^2 \phi_2)\cr
W & = {1 \over 27} ( 2 \del \phi_1 + \del \phi_2)(2 \del \phi_2 + \del
\phi_1)(\del \phi_1 - \del \phi_2) \cr
& \qquad - {1 \over 3} (\del \phi_1 \del^2 \phi_1 - \del \phi_2 \del^2
\phi_2) + {1 \over 6}
(\del \phi_1 \del^2 \phi_2 - \del \phi_2 \del^2 \phi_1) 
+ {1 \over 6} (\del^3 \phi_1 - \del^3 \phi_2 ) \cr
}$$
where $\phi_i = \alpha_i \cdot \phi$. There are
analogous quantities in the anti-holomorphic sector.

To construct conserved currents systematically in 
a general $\X_n$ Toda model, the equations of motion can be written 
as the zero-curvature condition for an auxiliary, two-dimensional gauge field 
with values in $\X_n$. From this gauge field one can then construct a 
Lax operator which in the simplest cases\note{This form for $\L$ holds 
for the algebras A, B, C, G which have no weight degeneracies in their
fundamental representations. For the remaining algebras there are some
additional factors of $\del^{-1}$ which make $\L$ a
pseudo-differential operator. Conserved quantities can be found in
the same way, however.}
can be written 
$$
\L(\X_n) = \prod_\lambda (\del + \lambda \cdot \del \phi) = 
\sum_r Y_r \, \del^r
\nfr{lax} 
where $\lambda$ are the weights of the fundamental representation of $\X_n$
taken in order of increasing height. 
The equations of motion imply $[\delb , \L(\X_n) ] = 0$, 
and the coefficients $Y_r$ written above are exactly the
conserved quantities which guarantee integrability of the model.
The spins of the conserved currents obtained in this way for the
models based on the classical algebras are
$$
\A_n : \, 2, 3, \ldots , n, n+1 \qquad 
\B_n \, \& \, \C_n : \, 2, 4, \ldots , 2n\qquad 
\D_n : \, 2, 4, \ldots , 2n-2, n
\nfr{spins}
For more details, see [\Ref{MS}] and references therein.

The pattern of spins above  
reflects the fact that it is possible to regard the Toda model
based on $\X_n$ as embedded in some sense in the Toda theory 
based on $\X_{n+1}$. 
Conversely, there is a precise way in which we can truncate the 
$\X_n$ Toda theory to the model one lower in the series based on
$\X_{n-1}$ (X = A, B, C or D).
This idea will prove important later, so we discuss how it works
in more detail.

To truncate the $\X_n$ Toda model to the $\X_{n-1}$ model we 
must discard one of the Toda fields, but we must
do this in a consistent
fashion taking due account of the exponential
interactions. 
First, let us agree to label the simple roots of the algebras 
of each type $\X$ in such a way that we can
regard the roots of $\X_{n-1}$ as a subset of those of $\X_n$. 
Now let $\omega$ be the highest weight of the fundamental 
representation of $\X_n$. Our choice of labelling means that 
$\omega \cdot \alpha_n = \alpha_n^2/2$ and $\omega \cdot \alpha_i =
0$ for $i \neq n$. 
Next set $\phi = - k \omega + \tilde \phi$ where $k$ is a constant and 
$\tilde \phi$
is defined to be orthogonal to $\omega$, ie.~to be a linear
combination of the simple roots of $\X_{n-1}$.
It is easy to check 
that if $\phi$ satisfies the $\X_n$ Toda equations, then 
$\tilde \phi$ satisfies the $\X_{n-1}$ Toda equations in the limit
$ k \rightarrow \infty$.
Intuitively this corresponds to taking the component of $\phi$
along $\omega$ to be equal to its classical vacuum value.

Now that we have a precise notion of how to truncate the Toda models
$\X_n \rightarrow \X_{n-1}$, let us see how this affects 
the conserved currents. 
Since the currents depend only on derivatives of
$\phi$, they are independent of $k$ and have well-defined
limits when we truncate by
taking $k \rightarrow \infty$. 
A conserved current of spin $q$ in the $\X_n$ model clearly 
descends to a conserved current 
of spin $q$ in the $\X_{n-1}$ model. 
It might become trivial in the truncated theory, of course, 
in the sense that it might be possible to write it as 
a combination of lower-spin conserved quantities. 
Indeed, this must happen when there is no independent conserved
quantity of spin $q$ in the truncated model. For example, the spin-4
current in
the $\A_3$ model must descend to a spin-4 current in the $\A_2$ model 
but, since there is no independent conserved quantity of spin-4, 
it can only be a combination of $T^2$, $\del^2 T$ and $\del W$. 

In fact it can be shown that 
all the currents 
occurring in the sequences given in \spins\ remain non-trivial under
truncation whenever possible, ie.~any current of spin $q$ will 
reduce to something non-trivial, 
provided there is an independent charge of spin $q$ in the truncated 
model.
This can be established by considering the effect of truncation on 
the Lax operator \lax .
Each $\L (\X_n)$ has a well-defined limit when we take $k \rightarrow
\infty$ and,
because $\omega$ is the highest weight of the fundamental representation, 
it is easy to relate this limit to $\L (\X_{n-1})$.
We find 
$\L(\A_n) \rightarrow \L(\A_{n-1}) \del$
and $\L(\X_n) \rightarrow \del \L(\X_{n-1}) \del$ for X = B, C or D.
It is therefore a relatively simple matter to keep
track of how the conserved quantities are related under truncation 
and so verify the claim. 
We will need this result later, specifically for the currents 
of spins 3 and 4.

\chapter{(1,0)-superconformal Toda models}

To write down a (1,0) supersymmetric extension of
the bosonic $\X_n$ Toda model we use 
(1,0) superspace. Let $\theta$ be a real fermionic coordinate 
transforming as a spinor of 
definite two-dimensional chirality, which will serve as a  
superpartner of $z$, and define the 
corresponding superderivative by $D = \del_\theta -i \theta \del$.
Consider the superspace Lagrangian 
$$
L = i D \Phi \cdot \bar \del \Phi +  \sum_a \left \{ \Psi_a D \Psi_a 
+ 2 \Psi_a \exp (\Half \alpha_a \cdot \Phi) \right \}
\nfr{sslag}
where the real scalar superfield $\Phi$ takes values in the Cartan
subalgebra of $\X_n$ and the fermions $\Psi_a$ 
$(a = 1, \ldots , n )$ are
Majorana spinors of the same chirality as $\theta$ 
(we write the $a$ index explicitly because 
we do not necessarily wish to identify the $n$-dimensional
space of fermions which it labels with the 
$n$-dimensional Cartan subalgebra).
To reduce this manifestly (1,0)-supersymmetric action to components we expand
the superfields
$$
\Phi = \phi + i \theta \lambda \ ,
\qquad 
\Psi_a = \psi_a + \theta \sigma_a \ ,
\nfr{comp}
and eliminate the scalar auxiliary fields $\sigma_a$ to arrive at the 
Lagrangian
$$
L = \del \phi \cdot \delb  \phi + i \lambda \cdot \delb \lambda 
+ \sum_a \left \{ i \psi_a \del \psi_a 
- \exp (\alpha_a \cdot \phi ) 
+ i (\alpha_a \cdot \lambda) \psi_a \exp (\Half \alpha_a \cdot \phi ) 
\right \}
\nfr{xslag}
with equations of motion 
$$\eqalign{
\del \delb \phi & = - \sum_a \Half \alpha_a \exp (\alpha_a \cdot \phi) 
+ \sum_a {\textstyle {1 \over 4}} i 
(\alpha_a \cdot \lambda) \psi_a \alpha_a \exp 
(\Half \alpha_a \cdot \phi) \ ,
\cr
\delb \lambda & = - \sum_a 
\Half \psi_a \alpha_a \exp (\Half \alpha_a \cdot \phi ) \ , \cr
\del \psi_a & = \Half (\alpha_a \cdot \lambda) 
\exp (\Half \alpha_a \cdot \phi ) \ . \cr
}
\nfr{eqm}
When the fermions $\lambda$ and $\psi_a$ are set to zero we recover
the action and equations of motion for the bosonic $\X_n$ Toda theory.
We shall refer to the (1,0)-supersymmetric extension as the 
$\S \X_n$ Toda model.

The $\S \X_n$ Toda theory 
is conformally-invariant, with the bosons transforming as in \conf\ and 
fermions transforming as 
$$ 
\lambda (z, \zb) \rightarrow (\del w)^{1/2} \lambda (w, \wb) \, , \qquad
\psi (z , \zb) \rightarrow (\delb \wb)^{1/2} \psi (w, \wb ) \ .
\nfr{fconf}
Notice that $\lambda$ and $\psi_a$ have opposite chirality and 
scaling dimensions $(1/2 , 0)$ and $(0 , 1/2)$ respectively.
In the holomorphic sector the conformal invariance is extended to
superconformal invariance 
by the 
supersymmetry transformations
$$
\delta \phi = i\epsilon \lambda \ , \qquad 
\delta \lambda = - \epsilon \del \phi \ , \qquad
\delta \psi_a = - \epsilon \exp (\Half \alpha_a \cdot \phi) \ ,
\nfr{susy}
where $\epsilon$ is a real spinor. 
Corresponding to these symmetries, we have spin-2 holomorphic and
anti-holomorphic components of the energy-momentum tensor
and a spin-3/2 holomorphic supersymmetry generator: 
$$\eqalign{
T & = \Half \del \phi \cdot \del \phi + {\textstyle {i \over 2}} 
\lambda \cdot \del \lambda
- \rho \cdot \del^2 \phi \, , \qquad 
\bar T = \Half \delb \phi \cdot \delb \phi + 
{\textstyle {i \over 2}} \psi_a \delb \psi_a
- \rho \cdot \delb^2 \phi \, ,\cr
G & = \Half \lambda \cdot \del \phi - \rho \cdot \del \lambda \, . \cr
}\efr 
It is easy to check using the equations of motion that these
are conserved. 

We now turn to the question of whether these (1,0) models admit an
additional supersymmetry of the opposite chirality, which would mean
that they were actually (1,1)-supersymmetric.
For this to be the case the fermions $\psi_a$ must be regarded as
living in the tangent bundle to the sigma-model target manifold.
The indices $a$ appearing on $\psi_a$ 
must then be taken as labeling a basis for the Cartan sub-algebra
of $\X_n$, and notice that this basis has already been assumed to be
{\it orthonormal\/}. 
In our case (a flat target manifold with no torsion)
the condition for (1,1) supersymmetry given in [\Ref{Pap},\Ref{HPT}]
is that the 
superpotential terms   
$\exp (\half \alpha_a \cdot \phi)$ 
should be the components 
of the exterior derivative of some function $f$ 
{\it in the orthonormal basis labeled by\/} $a$. 
Now it is natural to write these terms as functions of the 
coordinates 
$\phi_i = \alpha_i \cdot \phi$ on the target manifold, 
but this coordinate basis is 
{\it not\/} orthonormal.
By definition, the two bases are related by the matrix consisting 
of the components of the vectors $\alpha_i$ with respect to the 
orthonormal system:
$M_{a i} = (\alpha_i)_a$. This matrix 
satisfies $\sum_a M_{a i} M_{a j} =
\alpha_i \cdot \alpha_j$, which is the symmetrized Cartan matrix, and
we deduce that $M_{ai}$ is never diagonal unless it has rank one.
For (1,1) supersymmetry we now require the existence of a function
$f(\phi_i)$ for which 
$\del f / \del \phi_i = \sum_j M_{ij} \exp{ \half \phi_j}$. 
Clearly no such function exists unless 
we are dealing with the case of a single scalar field.
We conclude that the (1,0)-superconformal models are
not (1,1) supersymmetric, except in case based on $\A_1$ 
which gives the super-Liouville theory.
One can also check directly that there is no anti-holomorphic spin-3/2
quantity analogous to $G$ which could serve as the current 
corresponding to a second supersymmetry.\note{
It was claimed in [\Ref{Pap}] that all the (1,0) superconformal models
are actually (1,1) supersymmetric. The arguments above lead us to
disagree with this conclusion.
We also find that 
the additional supercharge proposed in eqn.~(21) of [\Ref{Pap}] is not
conserved.}

\chapter{Higher-spin quantities: low-rank superconformal models } 

We saw above that in each of the (1,0)-superconformal 
Toda theories there are
holomorphic and anti-holomorphic spin-2 components of the
energy-momentum 
tensor which differ from the bosonic expressions \bosem\ by 
modifications involving fermions.
We also saw that in the holomorphic sector the energy-momentum
tensor acquires a conserved spin-3/2 superpartner $G$.
Consider now how this might generalize to the higher-spin conserved
quantities present in the bosonic Toda models.

The simplest example is the spin-3 current $W$
in the $\A_2$ Toda theory. 
If the (1,0)-extension of this model is to be integrable, it is
natural to expect that 
$W$ will remain conserved
after a suitable modification by terms involving fermions, and it is
also natural to expect that it too 
will acquire a superpartner, this time of spin 5/2.
Assume for the moment that 
any such conserved currents 
must be polynomials in the fields $\del \phi$, $\lambda$ 
and their holomorphic derivatives. 
Then it is easy to see that the most general spin-5/2 quantity is a
linear combination of the terms
$$
\lambda_i \del \phi_j  \del \phi_k  , \quad
\lambda_i \del^2 \phi_j , \quad
\del \lambda_i \del \phi_j , \quad
\del^2 \lambda_i , \quad 
\lambda_i \lambda_j \del \lambda_k
$$
where $\phi_i = \alpha_i \cdot \phi$ and
$\lambda_i = \alpha_i \cdot \lambda$.
We can look similarly for the most general spin-3 quantity 
which could arise in modifying $W$, and
we find the candidate terms 
$$
\lambda_i \lambda_j \del \phi_k  \del \phi_l  , \quad
\lambda_i \lambda_j \del^2 \phi_k , \quad
\lambda_i \del \lambda_j \del \phi_k , \quad
\lambda_i \del^2 \lambda_j , \quad 
\del \lambda_i \del \lambda_j 
$$
After some lengthy computations, whose details we omit, we find that
even when the terms above are taken into account 
the only holomorphic 
quantity of spin 5/2 is $\del G$ and the only holomorphic quantity of spin 3
is $\del T$.
The conclusion is that the $\W$-symmetry of the bosonic $\A_2$ theory does
not survive in the $\S \A_2$ model. 

To clarify our assumptions concerning the nature of the conserved
currents,
consider the following list of quantities with definite
scaling-dimensions 
which arise in the general $\S \X_n$ Toda model:-- 
$$
\del \phi  \, : \, (1,0) \ , \quad
\delb \phi \, : \, (0,1) \ , \quad
\exp (\kappa \phi_i) \, : \, (\kappa , \kappa) \ , \quad 
\lambda \, : \, (\Half , 0)  \ , \quad
\psi_a \, : \, (0 , \Half)  \ .
\nfr{prim}
Our aim is to find the most general
holomorphic and anti-holomorphic currents of definite spin which can
be constructed as polynomials in the quantities above and in their
derivatives. We will not attempt to justify this starting point in any
more detail except to say that we consider it to be a rather weak 
condition on the composition of the currents.
We are interested in conserved quantities with
scaling-dimensions $(q,0)$ or $(0,q)$; but 
there are many ways of forming non-trivial
expressions such as 
$ \del \phi_i \delb \phi_j \exp (- \phi_k) $ or 
$ \lambda_i \psi_a \exp( - \half \phi_k)$ which have $h = \bar h = 0$
and which might therefore be expected to appear with arbitrary powers in the 
conserved quantities we seek.
The crucial point is that such terms can never appear as
part of a holomorphic or anti-holomorphic current. 

We claim that a holomorphic current of type $(q,0)$ can only arise as a
polynomial in quantities \prim\ and their derivatives which {\it all\/}
have $\bar h = 0$; similarly an anti-holomorphic current of type $(0,q)$
can only arise as a combination of these quantities which {\it all\/} 
have $h = 0$.
To see why this is true, consider the holomorphic case.
When we apply $\delb$ to any term involving $\phi$, $\psi_a$, or their
anti-holomorphic derivatives, the result cannot be simplified using
the equations of motion \eqm\ and the maximum power of $\delb$
appearing in any term is always increased. 
By contrast, when $\delb$ in applied to expressions involving only
$\del \phi$, $\lambda$ and their {\it holomorphic\/} derivatives, 
the results {\it can\/} be simplified using \eqm\ and there 
is a chance that they will conspire to cancel completely.
It is worth noting that these arguments apply just as well to the 
purely bosonic Toda models and that they provide one way of understanding
why only derivatives of $\phi$ appear in the Lax operators and 
conserved currents. 
This establishes our claim and justifies our assumption concerning the
composition of the 
holomorphic quantities which might appear in the the
$\S \A_2$ model. The same assumptions now apply to all 
the $\S \X_n$ models.

We carried out the calculations for the $\S \A_2$ model by hand, 
but to explore the
situation for other low-rank algebras it quickly becomes necessary to
use an algebraic manipulation package [\Ref{ALG}].
Proceeding in a similar way, we searched for all (anti-)holomorphic 
quantities with spins 
5/2 , 3, 7/2 or 4 for each of the classical algebras with rank $n \leq
4$ (there are nine independent cases).
For the reasons explained above, we assumed  
that the currents were polynomial in $\del \phi$, $\lambda$
($\delb \phi$, $\psi_a$) and their (anti-)holomorphic derivatives.
In each case, we found only those expressions that could be written
as combinations of $G$, $T$, $\bar T$ and their
derivatives---and as a useful check of the correctness of our
calculations we should note that we found {\it all\/} such
expressions. In the holomorphic sector, for example, 
we found only 
$$
\del G , \quad \del T , \quad GT , \quad \del^2 G , \quad
\del^2 T , \quad T^2 , \quad G \del G \, .
$$
We conclude that the bosonic $\W$-algebra structure is completely
destroyed in these supersymmetric theories. 

\chapter{Higher-spin quantities: general superconformal models} 

Having discussed some examples based on algebras of low rank,
let us see what can be inferred about the theories based on
general classical algebras. For definiteness we focus on 
the $\A_n$ series.
We contend that knowledge of the $\S \A_2$ case is enough to deduce that
the $\W$-algebra structure is spoiled in each of the 
$\S \A_n$ Toda models. More precisely, we will show that
there is no conserved spin-3 current in the $\S \A_n$ model which
reduces to the spin-3 current of the $\A_n$ model when the fermions are
set to zero.
 
We first generalize the truncation procedure of section 2 to the
(1,0)-superconformal models. 
Starting with the $\S \A_n$ theory, 
we proceed just as before for the bosonic fields
but in addition we discard superfluous fermions by
demanding that $\lambda$ lie in the space of roots of $\A_{n-1}$, 
setting $\psi_n = 0$, and checking that this is compatible with the
equations of motion \eqm . This defines a consistent truncation
of models $\S \A_n \rightarrow \S \A_{n-1}$. 

Suppose there were some spin-3 holomorphic current $\tilde W$ in the $\S
\A_n$ model which reduced to $W$, the spin-3 current of the
$\A_n$ model, when all fermions vanished. 
If we make the repeated truncation $\S \A_n \rightarrow \S \A_2$ 
then $\tilde W$ must still be a non-trivial holomorphic current, 
because we know from the arguments of section 2 that its 
purely-bosonic part $W$ remains non-trivial under the bosonic
truncation $\A_n \rightarrow \A_2$.
But we have already shown that there is no independent holomorphic 
spin-3 current 
in the $\S \A_2$ model. Hence there is no such current in the 
$\S \A_n$ model either. 
We can obviously deduce in a similar fashion 
that there is no independent holomorphic spin-4 current in the general 
$\S \A_n$ Toda model because we have shown that there is no 
such current in the $\S \A_3$ theory.

It is important to emphasize that these arguments forbid the existence 
of just those higher-spin currents which reduce to the bosonic Toda  
currents when the fermions vanish,
and knowledge of the bosonic case clearly plays a crucial role.
For instance, we cannot immediately deduce that there are no
spin-5/2 conserved quantities in the general $\S \A_n$ 
theory, despite having checked
explicitly that they are absent in
the $\S \A_2$, $\S \A_3$ and $\S \A_4$ models---we cannot be sure that
there is not a conserved spin-5/2 quantity in some higher-rank model
which becomes trivial on truncation to these theories. 
Having said that, it seems unlikely to us that such quantities exist,
since we have no reason to suspect that the higher-rank Toda theories
should behave qualitatively differently from the lower-rank ones for
which we have carried out explicit calculations.

The arguments used above for the $\A_n$ series 
can be extended immediately to the other
classical algebras. Based on the computations for the cases
of rank 4 and below we conclude that there are no generalizations of
the bosonic conserved currents
of spin 4 in any of the superconformal 
$\S \X_n$ models with X = B, C or D.

\chapter{Conclusions} 

We re-iterate our findings. The (1,0)-superconformal Toda models
do not admit (1,1) supersymmetry except for the case of the algebra
$\A_1$, corresponding to the super-Liouville theory.
Except for this simplest situation, the (1,0)-superconformal models 
do not contain generalizations of the conserved currents present in the
bosonic Toda theories.
We have checked that there are no independent conserved currents with spin 
5/2, 3, 7/2, 4 in any of the superconformal models based on the  
classical algebras with ranks 2, 3 or 4, 
and we have shown that this implies that there are no generalizations
of the bosonic conserved 
currents with spin 3 or 4 in the (1,0)-superconformal models
based on any of the classical algebras.

We think it is fair to interpret these results by saying that the
(1,0)-superconformal models are
not integrable in the usual sense of Toda theories. 
Of course, we cannot rule out the possibility 
that these models might
be integrable in some different sense. 
We have searched for conserved currents which
are either holomorphic or anti-holomorphic and which would therefore appear 
as part of some extended chiral algebra in the standard fashion
[\Ref{BS}]. 
However, it seems that some conformally-invariant field theories
possess conserved quantities with both holomorphic and
anti-holomorphic components [\Ref{AFGZ}] and one could even consider
the more general possibility of non-local conserved charges 
of the type familiar from work on non-linear sigma-models 
(see eg.~[\Ref{AAR}]). 
While these may be interesting issues for future study, we have no
grounds at present 
for suspecting that either of these types of conserved quantities are
to be found in the models considered here.

Our results confirm the conventional picture of 
how integrability of Toda theories with fermions seems to be linked
inextricably to  Lie superalgebras [\Ref{LLS}-\Ref{SHR}] and they
complement earlier work on the existence, or otherwise, of
various types of extended superconformal algebras
[\Ref{BS},\Ref{SW}].
They also provide a set of cautionary examples which
may challenge some preconceptions about the common properties of
field theories and their supersymmetric extensions.

We have not considered any exceptional algebras in our analysis.
There is no difficulty in principle in checking these models 
directly, following the approach of section 4. The problem is
that the complexity of the calculations increases 
very rapidly with both the number of fields and the spins of the 
conserved quantities being sought.
Perhaps some useful information about these models
could be gleaned from the results we already have, by some
suitable truncation to classical algebras. In any case, we see no
reason to anticipate any differences in the nature of the 
results that would be obtained.

Finally, the paper of Papadopoulos [\Ref{Pap}] also introduces
(1,0)-supersymmetric 
extensions of the affine Toda models corresponding to $\hat \A_n$. 
Based on the results obtained here, one might guess that
integrability fails for these models too, but it would be interesting
to investigate this further. 
One possibility would be to consider a modification of the truncation 
procedure discussed above in which the additional term in the
potential corresponding
to the affine root is scaled away to recover a conformal theory.
One might then be
able to deduce the absence of conserved charges from knowledge of the
conformal case.

\sjump
{\bf Acknowledgments:} JME is grateful to George Papadopoulos for
discussions and to the Theory Division, CERN for
warm hospitality while this work was being done. 
\bjump

\references

\beginref

\Rref{Pap}{G. Papadopoulos, Phys.~Lett.~{\bf B365} (1996) 98} 
\Rref{HPT}{C.M. Hull, G. Papadopoulos and P.K. Townsend,
Phys. Lett. {\bf B319} (1994) 291;\newline
G. Papadopoulos and P.K. Townsend, Class. Quantum Grav. {\bf 11}
(1994) 515; Class. Quantum Grav. {\bf 11} 2163}
\Rref{ALG}{E.S. Cheb-Terrab, 
{\sl Symbolic computing with grassmann variables}, preprint
IF-UERJ-95-27,
submitted to J. Symb. Comput., hep-th/9510226}

\Rref{LS}{A.N. Leznov and M.V. Saveliev, Lett. Math. Phys. {\bf3}
(1979) 489; Commun. Math. Phys. {\bf74} (1980) 11;
Commun. Math. Phys. {\bf 89} (1983) 59} 
\Rref{MS}{P. Mansfield and B. Spence, Nucl. Phys. {\bf B362} (1991) 294}
\Rref{O`R}{L. Feher, L. O'Raifeartaigh, P. Ruelle, I. Tsutsui and A. Wipf,
Phys. Rep. {\bf 222} (1992) 1}

\Rref{BS}{P. Bouwknegt and K. Schoutens, Phys. Rep. {\bf 223} (1993) 183}
\Rref{BW}{P. Bowcock and G.M.T. Watts, Nucl. Phys. {\bf B379} (1992) 63}
\Rref{SW}{T. Inami, Y. Matsuo and I. Yamanaka, Phys. Lett. {\bf B215}
(1988) 701;
K. Schoutens and A. Severin, Phys. Lett. {\bf B258} (1991) 134}

\Rref{LLS}{A.N. Leznov, M.V. Saveliev and D.A. Leites,
Phys. Lett. {\bf B96} (1980) 97}
\Rref{O}{M.A. Olshanetsky, Commun. Math. Phys. {\bf 88} (1983) 63} 
\Rref{LSS}{D.A. Leites, M.V. Saveliev and V.V. Serganova, `{\sl
Embeddings of Osp($N/2$) and associated non-linear supersymmetric
equations\/}, Proc.~Third Yurmala Seminar (USSR 22-24 May 1985),
{\sl Group theoretical methods in physics\/} vol.~1 (VNU Science Press,
Utrecht, 1986) }
\Rref{LM}{H.C. Liao and P. Mansfield, Nucl.~Phys.~{\bf B344} (1990)
696; Phys.~Lett.~{\bf B267} (1991) 188;
Phys.~Lett.~{\bf B252} (1991) 230; Phys.~Lett.~{\bf B252} (1991) 237
}
\Rref{EH}{J.M. Evans and T.J. Hollowood, Nucl. Phys. {\bf B352} (1991)
723, erratum Nucl. Phys. {\bf B382} (1992) 662;\newline
J.M. Evans, Nucl. Phys. {\bf B390} (1993) 225;\newline
J.M. Evans and T.J. Hollowood, Phys. Lett. {\bf B293} (1992) 100} 
\Rref{NM}{H. Nohara and K. Mohri, Nucl.~Phys.~{\bf B349} (1991) 253;
\newline
S. Komata, K. Mohri and H. Nohara, Nucl.~Phys.~{\bf B359} (1991) 168;
\newline
H. Nohara, Ann. Phys. {\bf 241} (1992) 1}
\Rref{SHR}{T. Inami and K-I. Izawa, Phys. Lett. {\bf B255} (1991) 521;
\newline
F. Delduc, E. Ragoucy and P. Sorba, Commun. Math. Phys. {\bf 146}
(1992) 403;
\newline
L. Frappat, E. Ragoucy and P. Sorba, Commun. Math, Phys. {\bf 157} 
(1993) 499}
\Rref{WI}{G.M.T. Watts, Nucl. Phys. {\bf B361} (1991) 311; \newline
K. Ito, Int. J. Mod. Phys. {\bf A7} (1992) 4885}
\Rref{IK}{T. Inami and H. Kanno, Commun. Math. Phys. {\bf 136} (1991)
519; Nucl. Phys. {\bf B359} (1991) 201}
\Rref{ABL}{C. Ahn, D. Bernard and A. LeClair, Nucl. Phys. {\bf B346}
(1990) 409} 
\Rref{GPZ}{S. Penati and D. Zanon, Nucl. Phys. {\bf B436} (1995) 265;
\newline
S. Penati, M. Pernici and D. Zanon, Phys. Lett. {\bf B309} (1993) 304}
\Rref{QI}{A. Gualzetti, S. Penati 
and D. Zanon, Nucl. Phys. {\bf B398} (1993) 622;
\newline
S. Penati and D. Zanon, Phys. Lett. {\bf B288} (1992) 297}
\Rref{STS}{M.T. Grisaru, S. Penati and D. Zanon, Nucl.~Phys.~{\bf B369} (1992)
373; Phys.~Lett.~{\bf B253} (1991) 357
\newline
G.W. Delius, M.T. Grisaru, S. Penati and D. Zanon, Nucl.~Phys.~{\bf
B359} (1991) 125; Phys.~Lett.~{\bf B256} (1991) 164
}
\Rref{AFGZ}{H. Aratyn, L.A. Ferreira, J.F. Gomes and A.H. Zimerman,
Mod. Phys. Lett. {\bf A9} (1994) 2783}
\Rref{AAR}{E. Abdalla, M.C.B. Abdalla and K.D. Rothe, 
{\sl Non-perturbative methods in 2 dimensional quantum field theory}
(1991) World-Scientific}
\endref
\ciao
